\documentclass[10pt,journal,letterpaper]{IEEEtran_nssmic}
\usepackage{graphicx}  
\usepackage{stfloats}  
\usepackage{amsmath}   
\begin{document}
\IEEEoverridecommandlockouts%
\title{Programmable Trigger Logic Unit Based on FPGA Technology}
%
\author{F.~Karstens\thanks{All authors are with the Universit\"at Freiburg, 79104 Freiburg, Germany.},~\IEEEmembership{Member,~IEEE,} S.~Trippel\thanks{Falk~Karstens is the corresponding author, falk.karstens@computer.org.}}%
%
\maketitle
\begin{abstract}
A programmable trigger logic module (TRILOMO) was implemented successfully in an FPGA using their internal look-up tables to save Boolean functions. Up to 16 trigger input signals can be combined logically for a fast trigger decision. 
The new feature is that the trigger decision is VME register based. 
The changes are made without modifying the FPGA code. 
Additionally the module has an excellent signal delay adjustment. 
\end{abstract}
\section{Introduction}
Increasing requirements on timing precision and speed of logic decisions in modern experiments demand the development of new fast and flexible logic modules. 
Field Programmable Gate Arrays (FPGA) allow programming of highly complex circuitries by combining a large number of small simple elements. 
They offer a high flexibility of the board after the hardware is build. 
The main idea is to program the FPGA using a Hardware Description Language (HDL) but keeping certain flexibility in the logic function. 
Distributed RAM can simulate any gate functionality in FPGAs. 
An VME interface is coded to select any logic function in a simple way. 
Thus a change in the logic function can be done on-the-fly without altering the HDL code. 
This has two advantages: first the final user of TRILOMO can change the logic without the knowledge of a sophisticated HDL and second the propagation time is independent of the logic function. 
That makes the presented module special against non processor FPGA trigger modules found elsewhere \cite{LeCroy}. 
Common applications for the board can be found in all experiments with triggered data acquisition. 

The basic principle of using RAM for the logic decision could also be achieved by commercial SRAM chips. 
The RAM in FPGAs is however optimized for very fast logic decisions and one gains in flexibility. 
Complex programmable logic devices (CPLD) and wired solutions miss the possibility to program different Boolean functions and to obtain high speed at the same time. 
Furthermore FPGAs are more cost effective for small scale projects than Application Specific Integrated Circuits (ASIC). 

A single slot 6U VME board has been designed and assembled with a FPGA as central element and NIM logic inputs/outputs. 
The functionality of the FPGA can be modified via the VME bus. 
The basic principle to use the RAM in the FPGA for fast and modifiable logic decisions is described in the next section. 
Additions required for full functionality in experimental environments like adjustable input delay and VME access are explained in section three. 
Section four describes a measurement of the time jitter of the leading edge of trigger signals passing the module. 
Applications of TRILOMO in experimental setups as well as new applications which are possible by reprogramming the FPGA are detailed in section five.

\section{Basic Concepts}
The VME board hosts 16 inputs and 1 output for the logic unit. 
The main processing happens in a central FPGA. 
The logic combination of the inputs is arbitrarily selectable. 
The result is given to the output in the fastest possible way. 
These main specifications make the board ideal to obtain fast trigger decisions. 

Look-up tables (LUT) are basic combinational elements of FPGAs and may be programmed as RAM. 
Spartan II XC2S200-6 \cite{Xi} produced by Xilinx Inc. provides blocks of 16 bit RAM with 0.5\,ns asynchronous read access. 
The use of RAM structure ensures constant propagation times independent of the
selected function. 
The logic functions of the LUTs are defined by the content of the RAM and can be easily modified by writing different values to the RAM. 
Since the RAM is accessable from a VME interface \cite{VME} the logic functionality of TRILOMO can be modified by the user without the knowledge of a HDL. 

\begin{figure}
\centering
\includegraphics[width=3in]{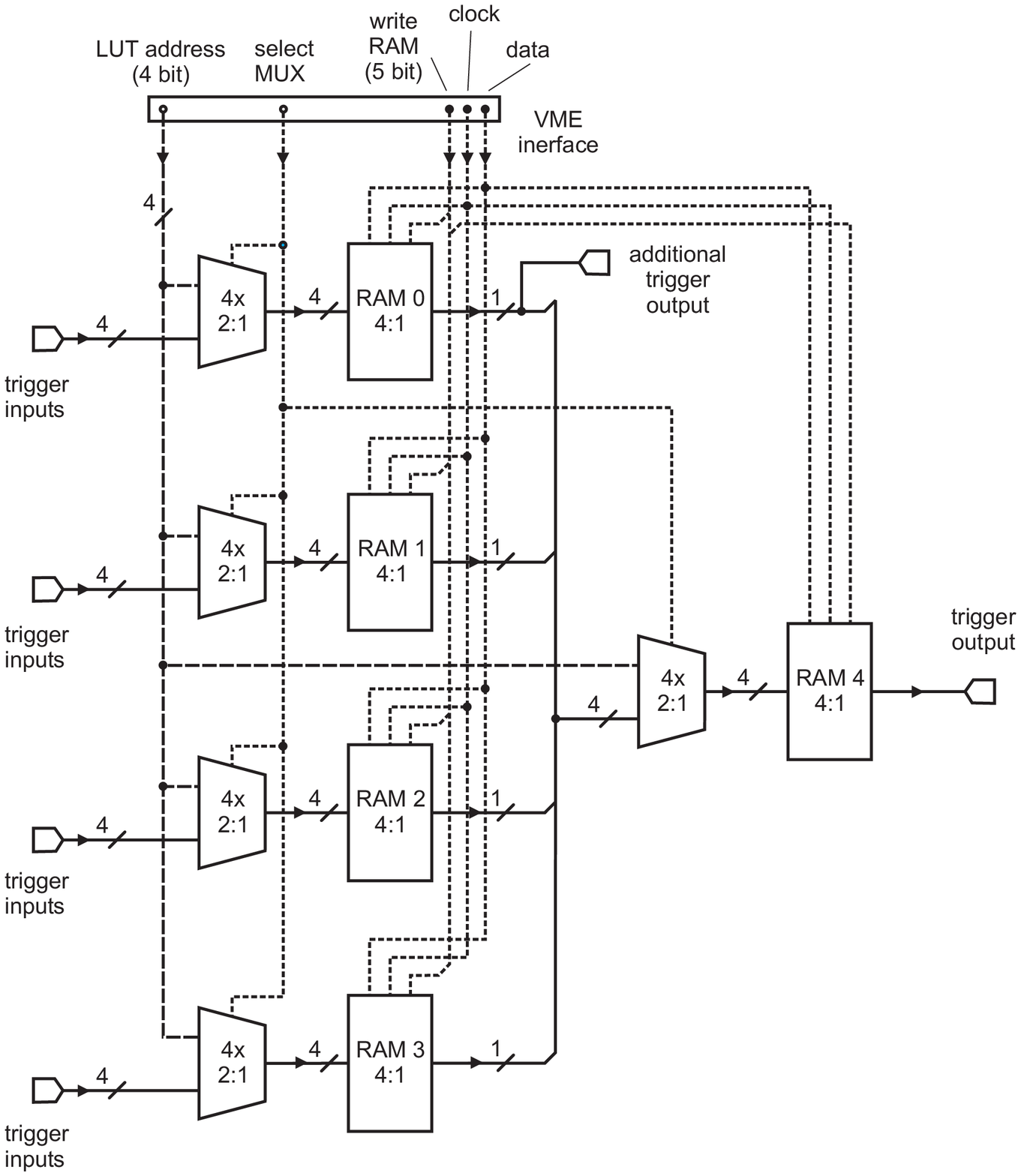}
\caption{Principle of wiring LUTs inside a FPGA. 
In standard mode 16 inputs select a saved bit what defines the output. 
In write mode all bits are redefined to select a new logic function which combines 16 inputs during read mode.}
\label{lut}
\end{figure}

The principle of wiring LUTs is shown in figure \ref{lut}. 
Four inputs are freely combined in a LUT in the first stage giving 16 possible signal combinations. 
In a second stage the output together with three other parallel outputs are combined again in a LUT to form the final trigger output. 
In standard mode the multiplexers pass the inputs to read the LUTs asynchronously (solid lines). 
In this mode the levels of the input lines form an address to read the memory. 
The stored bit of that address defines the output level. 

In write mode the multiplexer disables external input lines and instead enables lines to the VME bus to address the RAM (dashed lines). 
On a rising edge of a clock cycle a data bit is set for a given address of a
selected LUT while the RAM is hold in write access. 
Successively all 80 bits can be altered to define a Boolean function. 

In a mathematical subspace of four inputs all possible Boolean functions can be coded. 
In the full space of 16 inputs a large number of Boolean functions is possible but not the whole function space. 
This is due to the fact that one would need $2^{16}$ bits to implement all possible functions whereas in the presented solution $5\times 16=80$ bits are used for the trigger logic. 
However, any reasonable Boolean function should be implementable. 
%
\section{Logic Unit}
\begin{figure}
\centerline{\includegraphics[width=3in,angle=0]{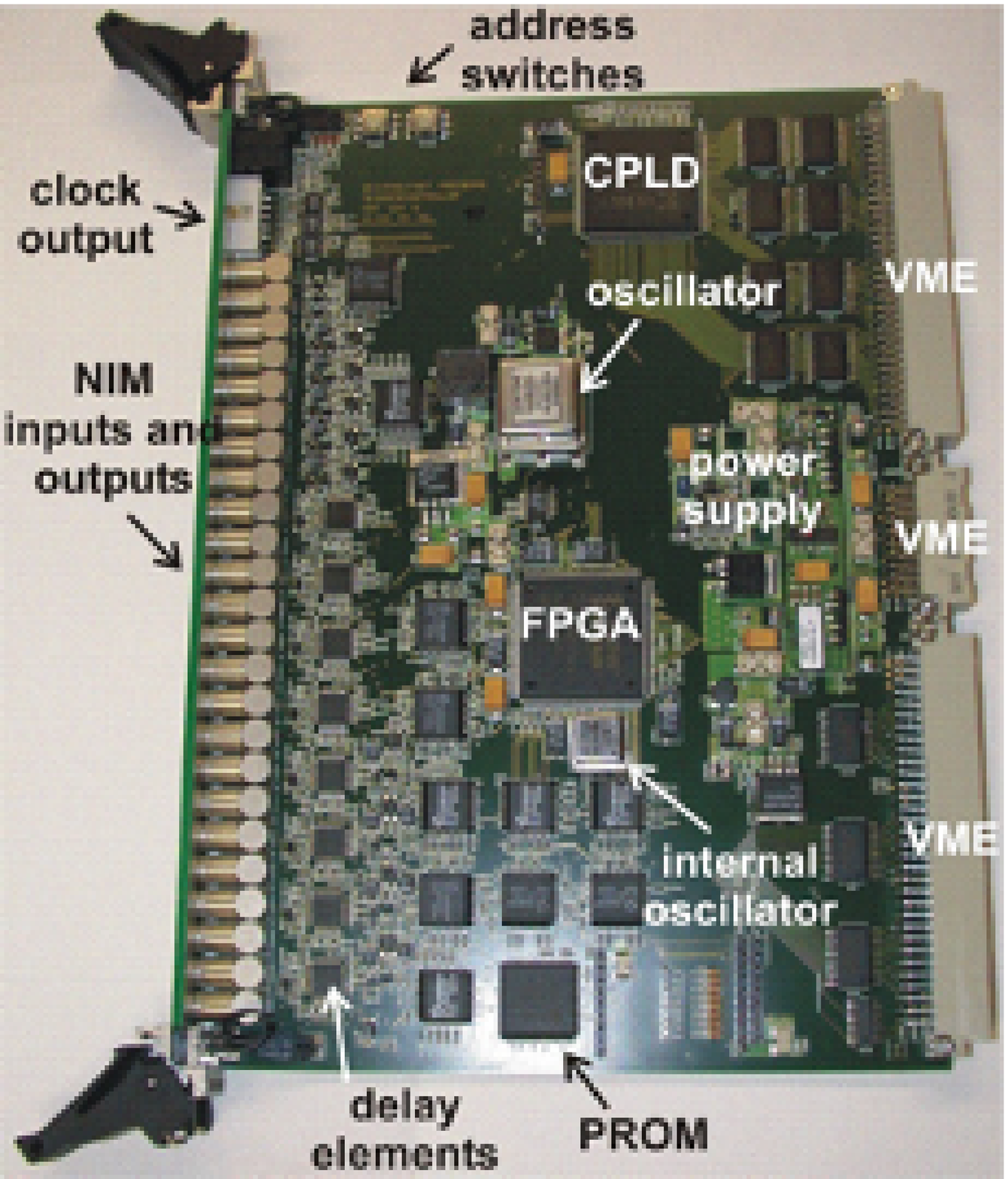}}
\caption{Picture of TRILOMO. The VME module has a size of 160\,mm $\times$ 233\,mm $\times$ 20\,mm (6U single slot).}
\label{photo}
\end{figure}

\begin{figure*}[!t]
\centerline{\includegraphics[width=6in,angle=0]{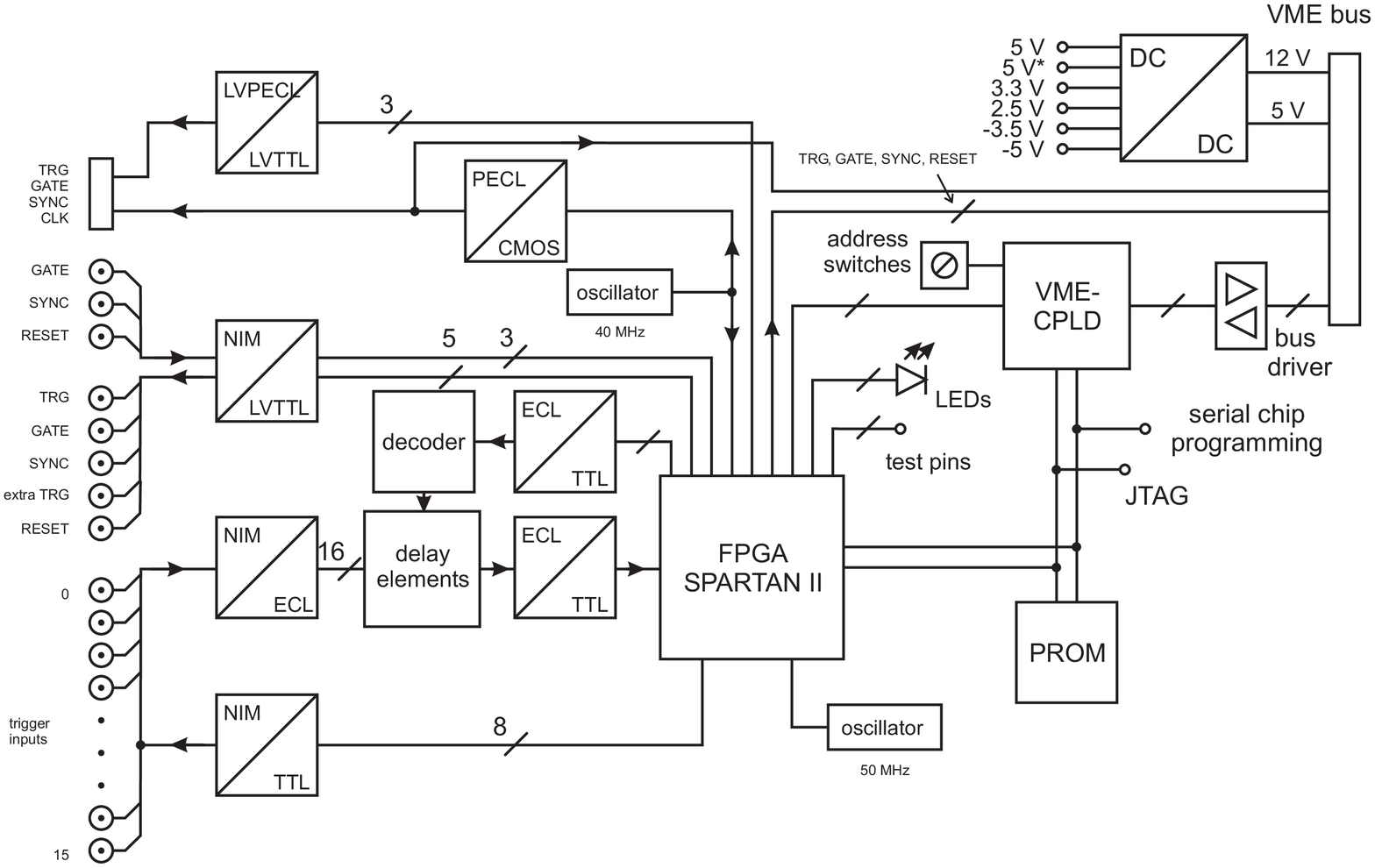}}
\caption{A block diagram showing how
the different chips are interconnected with each other (and the
number of lines). This helps the reader to gain a
better understanding of some design decisions. 
Inputs and outputs originating from the central FPGA are dominated by level converters and drivers. 
Additional main parts are the VME interface, delay elements with decoder, chip programming interface, power supply. 
Minor parts are test pins, LEDs, address switches and oscillators.
}
\label{blockdiagram}
\end{figure*}

A picture of the TRILOMO board is given in figure \ref{photo}. The adequate block diagram for a deeper understanding of the electronics structure is visible in figure \ref{blockdiagram}. 
Trigger inputs and outputs are current driven and follow the NIM standard (high input level -12\,mA to -32\,mA, output -14\,mA to -32\,mA; low level $>$ +2\,mA). 
The minimal input pulse width is 2\,ns. 
The output pulse width can be adjusted between 5\,ns and 30\,ns. 
The propagation time from any input to output is $(20\pm1)$\,ns. 
The time jitter of the output signal is less than 120\,ps with respect to the input signals. 

Different signal delays, either on board or outside, can be compensated by programmable delay lines which are implemented between the input plugs and FPGA. 
Consequently variation between inputs can be corrected to the order of a few picoseconds and hence coincidence times as short as 3.9\,ns are possible. 
For the delay we use single-channel chips of the type MC100EP195 \cite{PD}. 
The delay lines can be programmed individually via the VME interface in steps of 10\,ps up to 10.23\,ns. 

%

The key element, a Spartan II XC2S200-6,  
offers about 200,000 sytem gates and a system performance up to 200\,MHz. 
The FPGA can load the HDL design code (1.3\,Mbit) after power up via a PROM of type XC18V04 \cite{Xi}. 
Alternatively the program can be loaded via VME interface.  

A CPLD of the type X95288XL-10 \cite{Xi} realises VME access to the FPGA. 
It is connected through buffers to the 32-bit VME data bus and to the 32-bit VME address bus. 
Within the VME architecture the board acts as a VME slave module. 
The CPLD controls any write and read access between the FPGA and VME bus over a 16-bit address and a 32-bit data bus. 
By means of this bus the logic decision and other parameters can be redefined. 

Hot-swap circuitries \cite{LTC} are used such that the module can be safely inserted and extracted without the necessity to switch off the crate. 
The average power consumption is 15\,W. 
  
The eight layer board is optimized in terms of impedance matching and cross talk avoiding. 
The signal lines are in 200\,$\mu$m micro strip technology and the via size is 600\,$\mu$m. 

Additional gate, synchronisation, reset and clock inputs and outputs are implemented to integrate the module into a wider trigger electronics environment. Partially the signals are dublicated to backplane connectors or converted to ECL standard at the front of the module.  
%
%
%
\section{Jitter}
\begin{figure}
\centerline{\includegraphics[width=3in,angle=0]{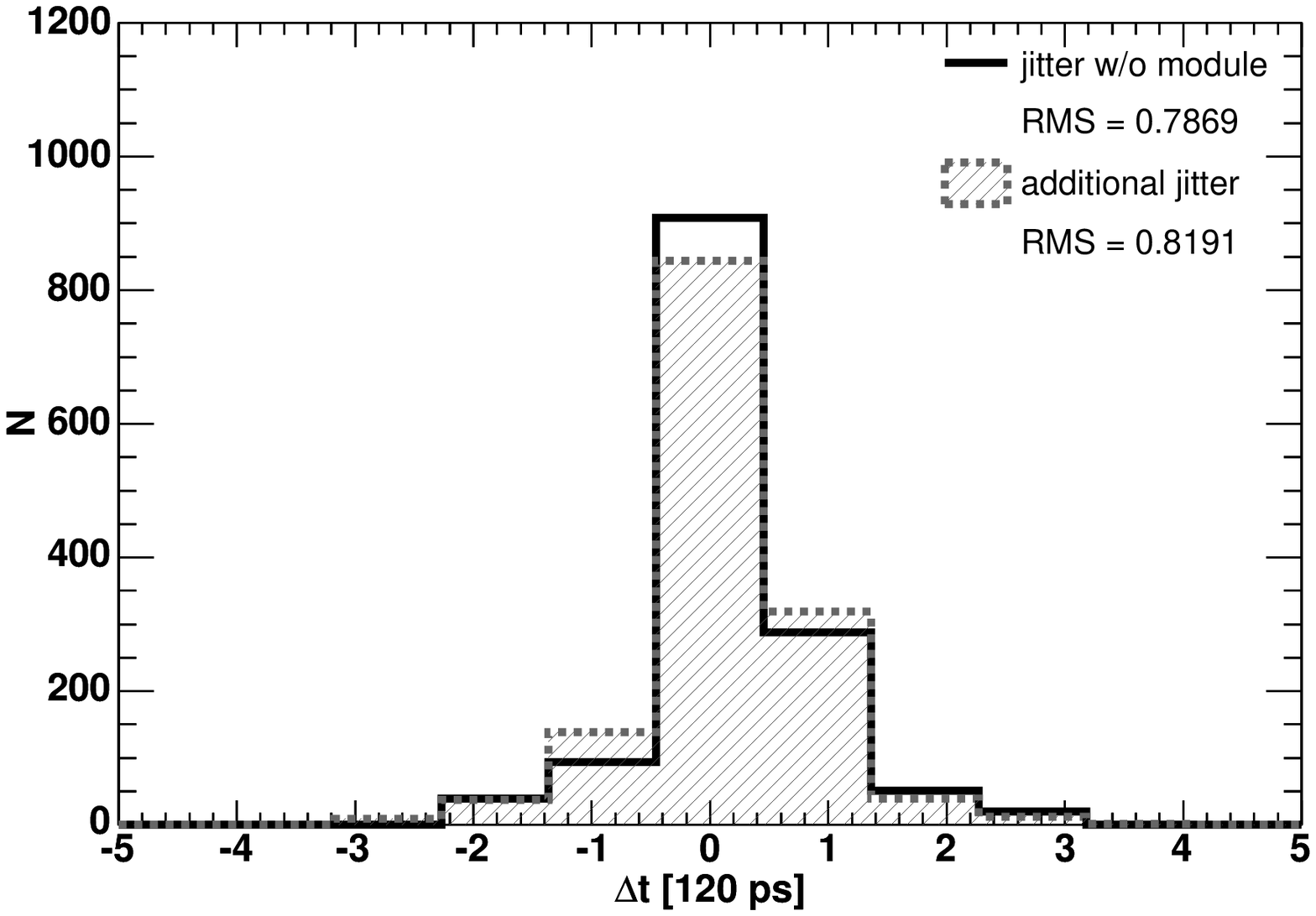}}
\caption{Time jitter measurement of the trigger logic module (dotted line). The underlying black curve reflects the resolution of the test setup. There is nearly no contribiution to the time jitter of the test setup beyond the time resolution of the test setup as can be seen in the comparison of RMS values.}
\label{jitter_plot}
\end{figure}

The time jitter of the leading edge of the input signal with respect to the output signal was measured beside functional tests and high frequency tests. 
The time jitter influences the minimal coincidence time given in a former section. 
Furthermore is the time jitter relevant if the time information of the leading edge of the trigger signal is used. 
A high resolution TDC \cite{F1} was used to measure the time jitter. 
The diagram in figure \ref{jitter_plot} shows the distribution of the time jitter. 
The underlying curve reflects the time resolution of the test setup, i.e. the TDC. 
The conclusion is that TRILOMO does not contribute additional time jitter above the time resolution of the test setup which is 120\,ps. 
%
%
\section{Applications}
TRILOMO is suitable for applications where fast trigger decisions are required. 
Typical applications are found in experiments of atomic, nuclear and particle physics. 
Short signal delays and precise timing allow high input rates up to 100\,MHz. 
The trigger decision can be set from a terminal on-the-fly without recabeling. 
The module can be cascaded so that the number of trigger inputs can be increased easily what makes experimental setups scalable. 

The number of applications of the module can be extended by the use of different HDL designs for the FPGA. 
Currently programs are available for counters, counting rates up to 150\,MHz, pre-scaler, I/O-registers and a complex veto trigger system.  
Other functions known from NIM modules can be easily implemented presuming knowledge of a HDL. 
Further applications are supported by the fact that eight inputs can be used as outputs. 
For precise timing applications an oven heated oscillator is mounted to the board. 
It delivers 40\,MHz with accuracy of 1\,ppm \cite{FO}. 
\section{Use at COMPASS}
TRILOMO is used in the NA58 experiment (COMPASS \cite{NA58}) at CERN as a veto counter to suppress halo particles of the muon beam. 
The module is connected to detectors sensitive to the sorrounding  beam area in front of a fixed target. 
The module blocks trigger signals which are time correlated with halo muons missing the target. 
High trigger rates demand a precise time correlation what the module is suitable for. 

The functionality of the module demanded by the experiment has been achieved with a rewritten VHDL program which adapts the module to an input register. 
High input rates up to 100\,MHz have been measured. 
\section*{Acknowledgment}
Special thanks are for the team -- the platform of the project.
The authors would like to thank the fruitful cooperation with the electronic workshop and mechanical workshop of the physics department of Freiburg University. 
We thank our colleagues from the front-end electronics group of the COMPASS collaboration at CERN for many useful discussions.  
This project has been supported by the German Bundesministerium f\"ur Bildung und Forschung.


\bibliographystyle{IEEEtranBST/IEEEtran.bst}
\bibliography{tns}

\end{document}